\documentclass[conference,a4paper]{APSIPA2020}
\usepackage{multirow}
\usepackage{graphicx}
\usepackage{amsmath}
\usepackage[psamsfonts]{amssymb}
\usepackage{amsxtra}
\usepackage{threeparttable}
\usepackage{cite}
\usepackage{grffile}
\usepackage{float}

\begin{document}

\title{Frequency Gating: Improved Convolutional Neural Networks for Speech Enhancement in the Time-Frequency Domain}

\author{%
\authorblockN{%
Koen Oostermeijer\authorrefmark{1}, Qing Wang\authorrefmark{1} and
Jun Du\authorrefmark{1}
}
\authorblockA{%
\authorrefmark{1}University of Science and Technology of China, Hefei, China  \\
E-mails: keen@mail.ustc.edu.cn, \{qingwang2, jundu\}@ustc.edu.cn}
}

\maketitle
\thispagestyle{empty}

\begin{abstract}
One of the strengths of traditional convolutional neural networks (CNNs) is their inherent translational invariance. However, for the task of speech enhancement in the time-frequency domain, this property cannot be fully exploited due to a lack of invariance in the frequency direction. In this paper we propose to remedy this inefficiency by introducing a method, which we call Frequency Gating, to compute multiplicative weights for the kernels of the CNN in order to make them frequency dependent. Several mechanisms are explored: temporal gating, in which weights are dependent on prior time frames, local gating, whose weights are generated based on a single time frame and the ones adjacent to it, and frequency-wise gating, where each kernel is assigned a weight independent of the input data. Experiments with an autoencoder neural network with skip connections show that both local and frequency-wise gating outperform the baseline and are therefore viable ways to improve CNN-based speech enhancement neural networks. In addition, a loss function based on the extended short-time objective intelligibility score (ESTOI) is introduced, which we show to outperform the standard mean squared error (MSE) loss function.
\end{abstract}
\section{Introduction}
The goal of speech enhancement is to clean an audio signal of an utterance from any corrupting background noise. In more formal terms, one seeks to increase the speech-to-noise ratio (SNR) to the point where the noise cannot be distinguished anymore and only the speech is heard.

Due to its applications in telecommunications and hearing aids, this has been an active area of research for many years. Classical speech enhancement algorithms include spectral subtraction methods \cite{Boll1979}, Wiener filtering \cite{Lim1979}, MMSE (minimum mean-square error) and OM-LSA (optimally modiﬁed log-spectral amplitude) estimation \cite{Ephraim1984, Cohen2001}, as well as subspace methods \cite{Dendrinos1991, Ephraim1995}.

Early applications of deep learning-based methods for speech enhancement employed fully connected neural networks (FNNs) \cite{Tamura1988, Xu2013, Liu2014, Xu2014} and recurrent neural networks (RNNs), in particular long short-term memory networks (LSTMs) \cite{Parveen2004, Weninger2015, Wollmer2013}.
Later, convolutional neural network-based methods (CNN) were introduced, first in the time-frequency domain \cite{Park2016, Fu2016, Kounovsky2017} and then in the time domain, for instance Denoising WaveNet \cite{Rethage2018}, an adaption of the eponymous WaveNet, and SEGAN \cite{Pascual2017, Baby2020, Phan2020}, which used a generative adversarial network (GAN) \cite{Goodfellow2014}.
\subsection{Related work}
In the time-frequency domain there are relatively few many-to-many frame (multiple time-frames as input and output) CNN-based methods, as most many-to-many methods operate in the time domain and most CNN-based methods in the time-frequency domain are many-to-one (single time-frame outputs).

Those that do make use of a many-to-many frame architecture include the following: In \cite{Tu2020}, the authors used a vanilla 24-layer CNN to estimate an ideal ratio mask (IRM). A similar idea is explored in \cite{Ribas2019}, in which a more sophisticated architecture with batch normalisation \cite{Ioffe2015} and skip connections \cite{He2016} was used to directly estimate LPS features  \cite{Du2008}. EHNet \cite{Zhao2018} employed a number of convolutional layers to generate a multiple feature maps, which are concatenated and fed into an RNN layer to better exploit temporal structures. The authors of \cite{Tolooshams2020} adapted U-net \cite{Ronneberger2015} for phase sensitive speech enhancement. They achieved this by constructing layers that are able to handle complex numbers. Finally, we mention \cite{Shah2018}, a GAN-based method.
\begin{figure*}[h]
\begin{center}
\includegraphics[width=0.8\textwidth]{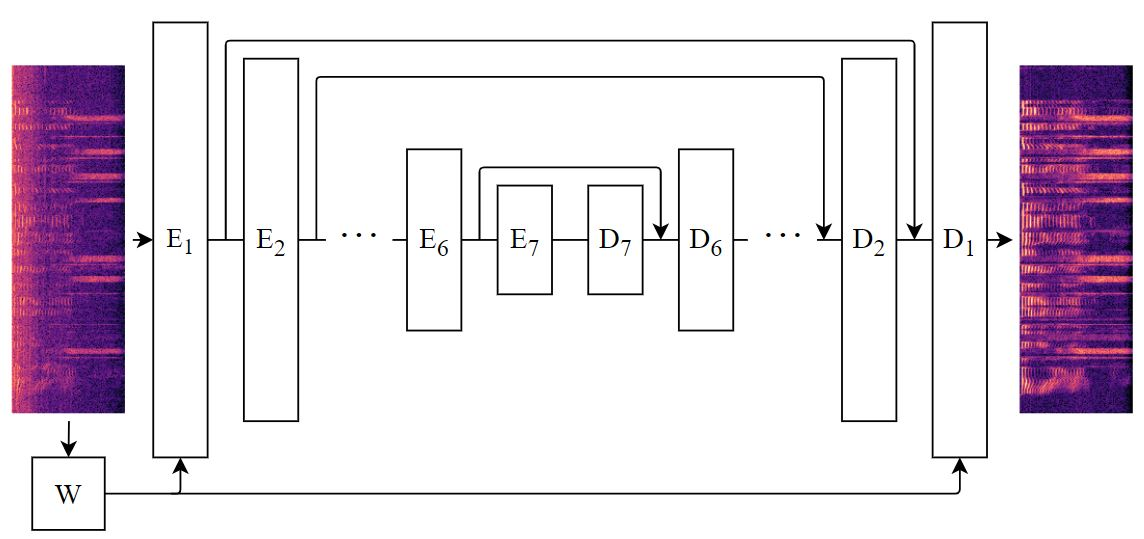}
\label{net}
\end{center}
\caption{Speech Enhancement Autoencoder. The components W, $\text{E}_i$ and $\text{D}_i$ denote the weighting, encoder and decoder layers respectively}
\vspace*{-3pt}
\end{figure*}
\subsection{Contributions}
Most of the time-frequency domain CNN architectures use kernels that span the whole frequency range in a sliding window fashion, or use large kernels with large strides in the frequency direction. We believe this to be a fundamental drawback, which we seek to amend by proposing a CNN autoencoder architecture with a frequency gating mechanism.

In chapter two we will substantiate this claim and go into detail on the reasoning behind our architecture. In addition to that, we also introduce a new loss function in chapter three. Chapter four is dedicated to describing our experiments, the results of which are presented in chapter five and discussed in chapter six.

To summarise, our main contributions are as follows:
\begin{enumerate}
    \item A new frequency gating mechanism is introduced to improve CNN-based speech enhancement in the time-frequency domain.
    \item We design a CNN-based autoencoder network for speech enhancement.
    \item We also introduce a new loss function, based on ESTOI.
    \item We show that these yield significantly better results over our baselines.
\end{enumerate}

\section{Denoising CNNs}
One of the hallmarks of CNNs is their inherent translational invariance; their ability to detect patterns is independent of where they are located in the input. This is beneficial for tasks such as image recognition and denoising, where it is assumed that features are not constrained to occur in certain regions of the image. However, for speech enhancement in the time-frequency domain this assumption cannot be expected to hold. While there is invariance in the temporal direction, patterns such as overtones in speech and the pinkness of noises break frequency invariance.

Previously, most networks dealt with this by having input kernels that have a height equal to the number of frequencies \cite{Park2016, Fu2016, Kounovsky2017}. However, such large kernels require a considerable amount of computation and thereby restrict the architecture and capabilities of the network. Consequently, these networks only estimate a single time-frame at a time.

In this paper, we investigate another option, instead of tall kernels, we use smaller kernels and introduce a frequency gating layer, tasked with computing frequency-dependent weights for the individual kernels. At each time step, this layer outputs for each kernel of the first and last convolutional layer and for each bin in the frequency direction a multiplicative weight between 0 and 1.

We explore three options of computing these weights: frequency bin weighting, local weighting and temporal weighting. Frequency bin weighting assigns kernel weights to each frequency, independent of the input. This explicitly breaks frequential invariance. The local and temporal weighting methods on the other hand generate a set of kernel weights per time frame, based on the whole frequency range of a single time frame and its neighbours, and all previous time frames respectively. This is explained in more detail in section IV.

The main network, tasked with denoising the utterance, takes a two-dimensional input of noisy log-power spectrum (LPS) features and outputs the corresponding enhanced LPS features of the same shape \cite{Du2008}.

It has an hourglass-shaped mirrored encoder-decoder structure with symmetric skip-connections, as illustrated in Fig. 1. Each of the layers consists of a convolutional layer, a batch normalisation layer \cite{Ioffe2015} and a ReLU non-linearity. The convolutional layers are transposed in the decoder and the output layer is without batch normalisation and non-linearity.

It has no pooling layers and is therefore fully connected in the CNN sense. Autoencoding CNNs enjoy several properties that are desirable for the task at hand: As opposed to recurrent layers, the calculations of the convolutional layers are completely parallelisable, increasing computational speed. Furthermore, the fact that the the network is many-to-many frame, means the network is better adapted at learning contextual information. The symmetric structure of the network allows for a straightforward use of skip connections, which has already been proved successful in raw audio-to-raw audio speech enhancement \cite{Pascual2017} as well as image denoising \cite{Ronneberger2015}. These skip connections increase the expressive power of the network by letting information bypass the bottleneck and ease the flow of gradients, thereby improving the learning capacity of the network.

\section{Loss Functions}
Speech enhancement algorithms are often trained using a mean squared error (MSE) loss function. However, it has been shown that in most cases this approach is suboptimal \cite{Kolbaek2020}. This is due to the fact that the quality metrics of interest are perceptual scores such as Perceptual Evaluation of Speech Quality (PESQ) \cite{Rec2001} and Short-Time Objective Intelligibility Measure (STOI) \cite{Taal2011}, which do not correlate perfectly with MSE. These metrics are sensitive to perceptually important effects such as differences in loudness, threshold effects and correlations, which the MSE loss fails to capture \cite{Martin2018}.

Since PESQ and STOI are calculated using discontinuous functions, they are unsuited to be used as loss functions directly. Several approaches have been proposed to bridge this gap. One popular method is to use a GAN \cite{Pascual2017}. The idea is that the discriminator is able to learn features of realistic speech and transfer this knowledge to the generator, which serves as the speech enhancement network. Another procedure is to augment the MSE loss function with symmetric and asymmetric loudness disturbance terms as a way to approximate the PESQ function \cite{Martin2018}. In \cite{Fu2019} the authors train a neural network to estimate PESQ values. Its weights are then frozen and used in the second training stage to optimise for PESQ indirectly.

\subsection{ESTOI}
In this paper, we propose a loss function based on ESTOI (Extended Short-Time Objective Intelligibility) \cite{Jensen2016}. This has been done previously in \cite{Kolbaek2020}, but for time-domain algorithms for which the data had been preprocessed using a voice activity detector. We have altered it to make it suitable for LPS features and more stable in situations with potentially a large proportion of silent frames.

ESTOI was introduced to remedy STOIs inability to capture the (un)intelligibility of speech distorted by temporally highly modulated noise. Like STOI, ESTOI is calculated using one-third octave frequency bands: let $S(k, m)$ be the short-time Fourier transform (STFT) of a signal $s(n)$ of which the silent frames have been removed. Here the indices $k = 1, \dots, K$ and $m = 1, \dots, M$ denote the frequency bin and the time frame respectively. The one-third octave bands are then obtained by summing over the bin energies:
\begin{align}
    S_j(m) = \sqrt{\sum_{k = k_l(j)}^{k_u(j)}|S(k, m)|^2},\label{octave}
\end{align}
where $k_l(j)$ and $k_u(j)$ are the lower and upper index of frequency band $j = 1, \dots, J$ respectively. These are cut up into segments $\{\bar{S}(N), \bar{S}(N+1), \dots, \bar{S}(M)\}$ of length N:
\begin{align}
    \bar{S}(m) = \begin{bmatrix} 
    S_1(m-N+1) & \dots &  S_1(m)\\
    \vdots & \ddots & \vdots\\
    S_J(m-N+1) & \dots  & S_J(m) 
    \end{bmatrix}\label{otob}
\end{align}
and subsequently normalised, first row-wise and then column-wise. Here, row-wise normalisation is defined for the $j$th row of $\bar{S}(m)$, $\bar{s}_j(m) = [S_1(m-N+1), \dots,  S_1(m)]^T$ as
\begin{align}
    \tilde{s}_j(m) = \frac{\bar{s}_j(m) - \mu_{\bar{s}_j(m)}}{||\bar{s}_j(m) - \mu_{\bar{s}_j(m)}||_2},\label{norm}
\end{align}
where $\mu_{\bar{s}_j(m)}$ is the average of the element of the vector $\bar{s}_j(m)$. Column-wise normalisation is done in a similar manner.

The row and column-wise normalised matrix of the clean speech $\tilde{C}(m)$ and of the enhanced speech $\tilde{N}(m)$ are then used to obtain the ESTOI measure:
\begin{align}
    d_\text{ESTOI} = \frac{1}{N(M-N+1)}\sum_{m=1}^{M-N+1} \text{Tr}\left(\tilde{C}(m)^T\tilde{N}(m)\right). \label{destoi}
\end{align}
Note that since this is a correlation function, it ranges from $-1$ to $1$, with $d_\text{ESTOI} = 1$ corresponding to perfect speech enhancement.
\begin{figure*}[t]
\begin{center}
\includegraphics[width=0.7\textwidth]{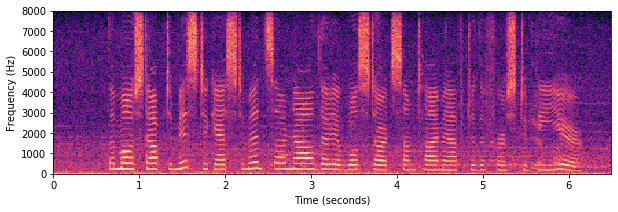}
\label{stft}
\end{center}
\caption{Spectrogram of an utterance corrupted by SNR = 5 noise.}
\vspace*{-3pt}
\end{figure*}

\begin{figure*}[t]
\begin{center}
\includegraphics[width=0.7\textwidth]{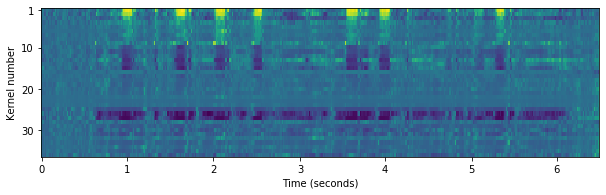}
\label{kernel}
\end{center}
\caption{Centred kernel activations, corresponding to Fig. 2.}
\vspace*{-3pt}
\end{figure*}
\subsection{$\text{E}^2$STOI}
To relieve the previously mentioned complications, we have extended ESTOI in three ways. First, during the beginning of training, transforming normalised LPS features back to absolute STFT features proved to be unstable. To ameliorate this we clip the data after it has been transformed into absolute STFT features to lie within the range $[0, 1]$.

Second, instead of removing silent frames from the dataset directly, silent frames are masked during the computation of the loss. Samples with too few non-silent frames remaining are then discarded. We found that this often left us with an insufficient number of remaining frames to justify splitting up the data into segments. Therefore, to improve stability we refrained from doing so. In the normalisation procedure only the remaining frames were considered.

Finally, we have added an MSE loss function for two reasons. Firstly because ESTOI is invariant under rescalings of the power spectra of the inputs and secondly our $\text{E}^2$STOI loss function only takes non-silent frames into account. The MSE loss function handles both these blind spots.

Hence, consider again an absolute power spectrum, clipped to the range $[0, 1]$. The mask is the set of frames for which the energy condition is fulfilled:
\begin{align}
    I = \left\{m \left|\sum_{k=1}^K H(|S(k, m)|) > \theta\right\}\right.,
\end{align}
here $H$ is the Heaviside step function and $\theta$ is the energy threshold, which was set to $\theta = 0.01$ in our experiments. The mask is used to filter the one-third octave frequency band features (Eq. \ref{otob}),
\begin{align}
    \underbar{S} = \begin{bmatrix} 
    S_1(l_1) & \dots &  S_1(l_L)\\
    \vdots & \ddots & \vdots\\
    S_J(l_1) & \dots  & S_J(l_L) 
    \end{bmatrix}, l_i \in I, l_i < l_j \forall i < j,
\end{align}
where $L = |I|$ is the cardinality of $I$, i.e. the number of non-silent frames.

Next the masked STFT power spectrum $\underline{S}$ is normalised in fashion similar to Eq. \ref{norm}, with a mean of
\begin{align}
    \mu_{\underline{s}_j(m)} = \frac{1}{L}\sum_{m \in I} S_j(m).
\end{align}
Then, $d_{\text{E}^2\text{STOI}}$ is computed  using STFT power spectra of the clean and the enhanced utterances, similar to Eq.\ref{destoi}. An MSE loss function is added to yield the final expression for our $\text{E}^2\text{STOI}$ loss function:
\begin{align}
    \mathcal{L}_{\text{E}^2\text{STOI}} = -d_{\text{E}^2\text{STOI}} + \lambda\mathcal{L}_\text{MSE}.
\end{align}
Experiments showed that for $\lambda = 1/3$ the magnitude of the gradients of both loss functions on the right-hand side are of equal magnitude.

\section{Experiments}
\subsection{Dataset}
The models are trained and evaluated on speech from the WSJ0 corpus \cite{WSJ0}, split into two groups of mutually exclusive speakers, resulting in a 13 hour training dataset and 42 minutes testing dataset. In constructing the training data samples, there is a 50\% overlap between consecutive samples.

The noise is taken from the DEMAND dataset \cite{demand}. For training we opted for the Metro, Square, Caf\'e, Station, Restaurant, Meeting, Hallway, Park, Field and Washing noises and testing was done using River, Cafeter and Traffic. In this paper, we have considered three SNR levels, -5 dB, 0 dB and 5 dB.

These waveforms were sampled at 16 kHz and short-time Fourier transformed using a 512 sample (32 msec) frame length and frame shift of 256 samples, resulting in 257 frequency bins, ranging from 0 to 8 kHz. From these, LPS features were extracted and normalised using speech only, as this gave slightly better results than to normalise it with noisy speech. They were split up into samples of length 40. We found that for this sample length a minimum number of non-silent frames of 10 worked well.

\subsection{Architectural Specifications}
The first and last three layers have a kernel size of $(5 \times 5)$ and all other layers $(3 \times 3)$. The first and last layers have a stride of one, whereas the other layers have a stride of two in the frequency direction and one in the time direction. For our inputs of height 257, the bottleneck representation has height 2. All convolutional layers are zero-padded to ensure the samples stay the same duration.

The number of channels of the encoder are 1, $\rho$, $\rho$, $\rho$, $2\rho$, $2\rho$, $3\rho$ and $4\rho$, which is reversed for the decoder, $\rho$ is a scale factor that we set to rescale the networks to make them the same size.

\subsubsection{Frequency bin weighting}
The frequency-wise bin weighting layer assigns each kernel of the input and output layers a frequency dependent weight $\sigma\left(\alpha_k x/257 + \beta_k\right)$ where $x$ ranges from 3 to 255, the frequency bin numbers the kernels are applied to. The parameters $\alpha_k$ and $\beta_k$ and $k$ denote the kernel index and $\sigma(\cdot)$ is the sigmoid function. Note that this only adds $2\rho$ extra parameters to the main network.

\subsubsection{Local weighting}
For the local weighting, a convolutional layer with kernels sized $(257 \times 3)$ is used. Each kernel assigns a weight to one of the kernels of the input and output layers, which is also transformed using a sigmoid function.

\subsubsection{Temporal weighting}
An LSTM layer is used to generate the weights for the temporal weighting layer. Its outputs are linearly transformed to lie within the range $[0, 1]$.\\

These versions of frequency gating are benchmarked against the previously discussed speech enhancement autoencoder without frequency gating. To give a fair comparison the benchmark and the one with frequency-wise weighting have a scale factor $\rho = 37$, giving them 732823 and 732897 parameters respectively, while local and temporal weighting use a smaller scale factor: $\rho = 36$, resulting in 721657 and 736345 parameters respectively.

Moreover, we have implemented 2D-RFCNN \cite{Tu2020} and an LSTM with 2048 hidden cells as further comparison.

\section{Results}
The results of our experiments are shown in Tables I, II and III, corresponding to River, Cafeteer and Traffic noise respectively.

\subsection{Efficacy Speech Enhancement CNN-based Autoencoder}
For all noise types and SNR levels every version of our CNN-based autoencoder architecture (bottom five rows in Tables I, II and III) yields significantly better results than the 2D-RFCNN and LSTM networks. For example, compared with the LSTM network, our CNN-based model without frequency gating achieved on average a 1.4 STOI and 0.15 PESQ improvement.

\subsection{Efficacy Frequency Gating Layer}
It can be seen that the frequency bin weighting method (``Freq.-wise") outperforms the benchmark (``No weight") on a majority of tasks (fifteen out of eighteen) and performs equally well on two of them. Especially on the PESQ metric tasks it provides all of the best results and on the traffic noise it performs best for every single task. Local weighting (``Local") yields best results for almost all (eight of nine) of the STOI metric tasks, but does not give better results for the PESQ tasks. Finally, temporal weighting (``Temporal") does not result in substantial improvements.

\subsection{Efficacy $\text{E}^2$STOI}
To provide a fair comparison, we have tested the frequency bin weighting method, which performed best among the frequency gating networks, with an MSE loss function (``Freq.-wise (MSE)"). By doing so, we can see the improvements of the $\text{E}^2$STOI loss function (``Freq.-wise") over the standard MSE loss function. The experiments show that for the vast majority (seventeen out of eighteen tasks) the $\text{E}^2$STOI loss function is the better loss function.

\subsection{Weighting Layer}
To demonstrate how the frequency gating principle works, we have taken an SNR = 5 utterance and applied the trained local weighting layer to it. Fig. 2 shows the utterance and Fig. 3 centred kernel weights. The first nine kernels are activated by the eight high-energy patterns in the high frequency regions, whereas the next six kernels are deactivated here. Kernels 17 to 24 are always activated, regardless of any patterns in the input. Furthermore, we can see several kernels that deactivate when there is speech activity.

Moreover, inspecting the $\alpha_k$ parameters of the frequency bin weighting layer shows that 21 of those are positive while 16 are negative. This means that the majority of the kernel weights get decrease the higher the frequency.

\begin{table}[H]
\begin{center}
\label{river}
\caption{River noise; STOI(\%) and PESQ comparison}
\begin{tabular}{l|cccccc}
 & \multicolumn{2}{c}{SNR = -5} & \multicolumn{2}{c}{SNR = 0} & \multicolumn{2}{c}{SNR = 5} \\
 & STOI & PESQ & STOI & PESQ & STOI & PESQ \\
 \hline
Noisy & 63.6 & 1.60 & 78.5 & 1.91 & 88.3 & 2.26\\
2D-RFCNN & 64.2 & 1.63 & 79.5 & 2.00 & 89.4 & 2.37\\
LSTM & 67.8 & 1.73 & 80.3 & 2.06 & 89.2 & 2.44\\
No weight & 67.7 & 1.80 & 81.4 & 2.15 & 90.2 & 2.53 \\
Temporal & 68.8 & 1.72 & 82.2 & 2.07 & 90.2 & 2.43 \\
Local & \textbf{69.4} & 1.77 & \textbf{82.4} & 2.11 & \textbf{90.5} & 2.47 \\
Freq.-wise & 68.3 & \textbf{1.81} & 81.9 & \textbf{2.16} & 90.4 & \textbf{2.55} \\
Freq.-wise (MSE) & 66.3 & 1.71 & 80.5 & 2.06 & 89.5 & 2.46
\end{tabular}
\end{center}
\end{table}

\begin{table}[H]
\label{cafeteer}
\caption{Cafeteer noise; STOI(\%) and PESQ comparison}
\begin{tabular}{l|cccccc}
 & \multicolumn{2}{c}{SNR = -5} & \multicolumn{2}{c}{SNR = 0} & \multicolumn{2}{c}{SNR = 5} \\
 & STOI & PESQ & STOI & PESQ & STOI & PESQ \\
 \hline
Noisy & 50.5 & 1.54 & 68.6 & 1.87 & 83.0 & 2.22\\
2D-RFCNN & 52.3 & 1.58 & 70.1 & 1.93 & 84.5 & 2.31 \\
LSTM & 53.9 & 1.65 & 71.3 & 2.00 & 84.7 & 2.37\\
No weight & 54.1 & 1.75 & 72.7 & \textbf{2.11} & 86.0 & 2.47 \\
Temporal & 54.6 & 1.72 & 72.8 & 2.08 & 85.8 & 2.44 \\
Local & \textbf{55.3} & 1.74 & \textbf{73.5} & 2.10 & \textbf{86.4} & 2.46 \\
Freq.-wise & 54.2 & \textbf{1.76} & 72.6 & \textbf{2.11} & 86.1 & \textbf{2.49} \\
Freq.-wise (MSE) & 54.6 & 1.74 & 73.1 & 2.10 & 86.1 & 2.47
\end{tabular}
\end{table}

\begin{table}[H]
\label{traffic}
\caption{Traffic noise; STOI(\%) and PESQ comparison}
\begin{tabular}{l|cccccc}
 & \multicolumn{2}{c}{SNR = -5} & \multicolumn{2}{c}{SNR = 0} & \multicolumn{2}{c}{SNR = 5} \\
 & STOI & PESQ & STOI & PESQ & STOI & PESQ \\
 \hline
Noisy & 71.6 & 1.78 & 84.0 & 2.15 & 92.0 & 2.54\\
2D-RFCNN & 73.2 & 1.87 & 85.3 & 2.26 & 92.7 & 2.65 \\
LSTM & 76.2 & 1.98 & 86.0 & 2.37 & 92.1 & 2.74\\
No weight & 79.4 & 2.18 & \textbf{89.1} & 2.59 & 94.7 & 2.97 \\
Temporal & 78.9 & 2.07 & 88.7 & 2.48 & 94.5 & 2.86 \\
Local & \textbf{79.6} & 2.14 & \textbf{89.1} & 2.55 & 94.7 & 2.93 \\
Freq.-wise & \textbf{79.6} & \textbf{2.20} & \textbf{89.1} & \textbf{2.60} & \textbf{94.9} & \textbf{3.00}
 \\
Freq.-wise (MSE) & 79.3 & 2.09 & 88.8 & 2.51 & 94.3 & 2.90
\end{tabular}
\end{table}

\section{Conclusion and Discussion}
Our CNN-based autoencoder design proved to be very effective, outperforming our baselines by a large margin. Furthermore, it was shown that frequency gating can significantly improve CNN-based speech enhancement networks at only a small cost. Frequency-wise weighting resulted in the best overall improvement, while only adding a relatively small amount of parameters to the network, an increase of about $10^{-4}$\% in the number of parameters in our case. Local weighting gave the best improvements for STOI, but did not show as many improvements compared to the benchmark when it came to PESQ. We hypothesise that this is at least partly due to the fact that optimised for $\text{E}^2$STOI, which is closely related to STOI. A loss function more strongly correlated with PESQ could yield better results. Furthermore, it was shown that the weights do indeed serve their intended purpose and adapted to different patterns in the frequency range. Temporal weighting, however, did not show any improvement. Finally, the $\text{E}^2$STOI loss function also yielded significant improvements.


\bibliographystyle{unsrt}
\bibliography{bib.bib}

\begin{thebibliography}{10}

\bibitem{Boll1979}
Steven Boll.
\newblock Suppression of acoustic noise in speech using spectral subtraction.
\newblock {\em IEEE Transactions on acoustics, speech, and signal processing},
  27(2):113--120, 1979.

\bibitem{Lim1979}
Jae~Soo Lim and Alan~V Oppenheim.
\newblock Enhancement and bandwidth compression of noisy speech.
\newblock {\em Proceedings of the IEEE}, 67(12):1586--1604, 1979.

\bibitem{Ephraim1984}
Yariv Ephraim and David Malah.
\newblock Speech enhancement using a minimum-mean square error short-time
  spectral amplitude estimator.
\newblock {\em IEEE Transactions on acoustics, speech, and signal processing},
  32(6):1109--1121, 1984.

\bibitem{Cohen2001}
Israel Cohen and Baruch Berdugo.
\newblock Speech enhancement for non-stationary noise environments.
\newblock {\em Signal processing}, 81(11):2403--2418, 2001.

\bibitem{Dendrinos1991}
Markos Dendrinos, Stelios Bakamidis, and George Carayannis.
\newblock Speech enhancement from noise: A regenerative approach.
\newblock {\em Speech Communication}, 10(1):45--57, 1991.

\bibitem{Ephraim1995}
Yariv Ephraim and Harry~L Van~Trees.
\newblock A signal subspace approach for speech enhancement.
\newblock {\em IEEE Transactions on speech and audio processing},
  3(4):251--266, 1995.

\bibitem{Tamura1988}
Shinichi Tamura and Alex Waibel.
\newblock Noise reduction using connectionist models.
\newblock In {\em ICASSP-88., International Conference on Acoustics, Speech,
  and Signal Processing}, pages 553--554, 1988.

\bibitem{Xu2013}
Yong Xu, Jun Du, Li-Rong Dai, and Chin-Hui Lee.
\newblock An experimental study on speech enhancement based on deep neural
  networks.
\newblock {\em IEEE Signal processing letters}, 21(1):65--68, 2013.

\bibitem{Liu2014}
Ding Liu, Paris Smaragdis, and Minje Kim.
\newblock Experiments on deep learning for speech denoising.
\newblock In {\em Fifteenth Annual Conference of the International Speech
  Communication Association}, 2014.

\bibitem{Xu2014}
Yong Xu, Jun Du, Li-Rong Dai, and Chin-Hui Lee.
\newblock A regression approach to speech enhancement based on deep neural
  networks.
\newblock {\em IEEE/ACM Transactions on Audio, Speech, and Language
  Processing}, 23(1):7--19, 2014.

\bibitem{Parveen2004}
Shahla Parveen and Phil Green.
\newblock Speech enhancement with missing data techniques using recurrent
  neural networks.
\newblock In {\em 2004 IEEE International Conference on Acoustics, Speech, and
  Signal Processing}, volume~1, pages I--733. IEEE, 2004.

\bibitem{Weninger2015}
Felix Weninger, Hakan Erdogan, Shinji Watanabe, Emmanuel Vincent, Jonathan
  Le~Roux, John~R Hershey, and Bj{\"o}rn Schuller.
\newblock Speech enhancement with lstm recurrent neural networks and its
  application to noise-robust asr.
\newblock In {\em International Conference on Latent Variable Analysis and
  Signal Separation}, pages 91--99. Springer, 2015.

\bibitem{Wollmer2013}
Martin W{\"o}llmer, Zixing Zhang, Felix Weninger, Bj{\"o}rn Schuller, and
  Gerhard Rigoll.
\newblock Feature enhancement by bidirectional lstm networks for conversational
  speech recognition in highly non-stationary noise.
\newblock In {\em 2013 IEEE International Conference on Acoustics, Speech and
  Signal Processing}, pages 6822--6826. IEEE, 2013.

\bibitem{Park2016}
Se~Rim Park and Jinwon Lee.
\newblock A fully convolutional neural network for speech enhancement.
\newblock {\em arXiv preprint arXiv:1609.07132}, 2016.

\bibitem{Fu2016}
Szu-Wei Fu, Yu~Tsao, and Xugang Lu.
\newblock Snr-aware convolutional neural network modeling for speech
  enhancement.
\newblock In {\em Interspeech}, pages 3768--3772, 2016.

\bibitem{Kounovsky2017}
Tomas Kounovsky and Jiri Malek.
\newblock Single channel speech enhancement using convolutional neural network.
\newblock In {\em 2017 IEEE International Workshop of Electronics, Control,
  Measurement, Signals and their Application to Mechatronics (ECMSM)}, pages
  1--5. IEEE, 2017.

\bibitem{Rethage2018}
Dario Rethage, Jordi Pons, and Xavier Serra.
\newblock A wavenet for speech denoising.
\newblock In {\em 2018 IEEE International Conference on Acoustics, Speech and
  Signal Processing (ICASSP)}, pages 5069--5073. IEEE, 2018.

\bibitem{Pascual2017}
Santiago Pascual, Antonio Bonafonte, and Joan Serra.
\newblock Segan: Speech enhancement generative adversarial network.
\newblock {\em arXiv preprint arXiv:1703.09452}, 2017.

\bibitem{Baby2020}
Deepak Baby.
\newblock isegan: Improved speech enhancement generative adversarial networks.
\newblock {\em arXiv preprint arXiv:2002.08796}, 2020.

\bibitem{Phan2020}
Huy Phan, Ian~V McLoughlin, Lam Pham, Oliver~Y Ch{\'e}n, Philipp Koch, Maarten
  De~Vos, and Alfred Mertins.
\newblock Improving gans for speech enhancement.
\newblock {\em arXiv preprint arXiv:2001.05532}, 2020.

\bibitem{Goodfellow2014}
Ian Goodfellow, Jean Pouget-Abadie, Mehdi Mirza, Bing Xu, David Warde-Farley,
  Sherjil Ozair, Aaron Courville, and Yoshua Bengio.
\newblock Generative adversarial nets.
\newblock In {\em Advances in neural information processing systems}, pages
  2672--2680, 2014.

\bibitem{Tu2020}
Yan-Hui Tu, Jun Du, and Chin-Hui Lee.
\newblock 2d-to-2d mask estimation for speech enhancement based on fully
  convolutional neural network.
\newblock In {\em ICASSP 2020-2020 IEEE International Conference on Acoustics,
  Speech and Signal Processing (ICASSP)}, pages 6664--6668. IEEE, 2020.

\bibitem{Ribas2019}
Dayana Ribas, Jorge Llombart, Antonio Miguel, and Luis Vicente.
\newblock Deep speech enhancement for reverberated and noisy signals using wide
  residual networks.
\newblock {\em arXiv preprint arXiv:1901.00660}, 2019.

\bibitem{Ioffe2015}
Sergey Ioffe and Christian Szegedy.
\newblock Batch normalization: Accelerating deep network training by reducing
  internal covariate shift.
\newblock {\em arXiv preprint arXiv:1502.03167}, 2015.

\bibitem{He2016}
Kaiming He, Xiangyu Zhang, Shaoqing Ren, and Jian Sun.
\newblock Deep residual learning for image recognition.
\newblock In {\em Proceedings of the IEEE conference on computer vision and
  pattern recognition}, pages 770--778, 2016.

\bibitem{Du2008}
Jun Du and Qiang Huo.
\newblock A speech enhancement approach using piecewise linear approximation of
  an explicit model of environmental distortions.
\newblock In {\em Ninth annual conference of the international speech
  communication association}, 2008.

\bibitem{Zhao2018}
Han Zhao, Shuayb Zarar, Ivan Tashev, and Chin-Hui Lee.
\newblock Convolutional-recurrent neural networks for speech enhancement.
\newblock In {\em 2018 IEEE International Conference on Acoustics, Speech and
  Signal Processing (ICASSP)}, pages 2401--2405. IEEE, 2018.

\bibitem{Tolooshams2020}
Bahareh Tolooshams, Ritwik Giri, Andrew~H Song, Umut Isik, and Arvindh
  Krishnaswamy.
\newblock Channel-attention dense u-net for multichannel speech enhancement.
\newblock In {\em ICASSP 2020-2020 IEEE International Conference on Acoustics,
  Speech and Signal Processing (ICASSP)}, pages 836--840. IEEE, 2020.

\bibitem{Ronneberger2015}
Olaf Ronneberger, Philipp Fischer, and Thomas Brox.
\newblock U-net: Convolutional networks for biomedical image segmentation.
\newblock In {\em International Conference on Medical image computing and
  computer-assisted intervention}, pages 234--241. Springer, 2015.

\bibitem{Shah2018}
Neil Shah, Hemant~A Patil, and Meet~H Soni.
\newblock Time-frequency mask-based speech enhancement using convolutional
  generative adversarial network.
\newblock In {\em 2018 Asia-Pacific Signal and Information Processing
  Association Annual Summit and Conference (APSIPA ASC)}, pages 1246--1251.
  IEEE, 2018.

\bibitem{Kolbaek2020}
Morten Kolb{\ae}k, Zheng-Hua Tan, S{\o}ren~Holdt Jensen, and Jesper Jensen.
\newblock On loss functions for supervised monaural time-domain speech
  enhancement.
\newblock {\em IEEE/ACM Transactions on Audio, Speech, and Language
  Processing}, 28:825--838, 2020.

\bibitem{Rec2001}
ITU-T Recommendation.
\newblock Perceptual evaluation of speech quality (pesq): An objective method
  for end-to-end speech quality assessment of narrow-band telephone networks
  and speech codecs.
\newblock {\em Rec. ITU-T P. 862}, 2001.

\bibitem{Taal2011}
Cees~H Taal, Richard~C Hendriks, Richard Heusdens, and Jesper Jensen.
\newblock An algorithm for intelligibility prediction of time--frequency
  weighted noisy speech.
\newblock {\em IEEE Transactions on Audio, Speech, and Language Processing},
  19(7):2125--2136, 2011.

\bibitem{Martin2018}
Juan~Manuel Mart{\'\i}n-Do{\~n}as, Angel~Manuel Gomez, Jose~A Gonzalez, and
  Antonio~M Peinado.
\newblock A deep learning loss function based on the perceptual evaluation of
  the speech quality.
\newblock {\em IEEE Signal processing letters}, 25(11):1680--1684, 2018.

\bibitem{Fu2019}
Szu-Wei Fu, Chien-Feng Liao, and Yu~Tsao.
\newblock Learning with learned loss function: Speech enhancement with
  quality-net to improve perceptual evaluation of speech quality.
\newblock {\em IEEE Signal Processing Letters}, 27:26--30, 2019.

\bibitem{Jensen2016}
Jesper Jensen and Cees~H Taal.
\newblock An algorithm for predicting the intelligibility of speech masked by
  modulated noise maskers.
\newblock {\em IEEE/ACM Transactions on Audio, Speech, and Language
  Processing}, 24(11):2009--2022, 2016.

\bibitem{WSJ0}
John Garofalo, David Graff, Doug Paul, and David Pallett.
\newblock Csr-i (wsj0) complete.
\newblock {\em Linguistic Data Consortium, Philadelphia}, 2007.

\bibitem{demand}
Joachim Thiemann, Nobutaka Ito, and Emmanuel Vincent.
\newblock {DEMAND: a collection of multi-channel recordings of acoustic noise
  in diverse environments}, June 2013.
\newblock {Supported by Inria under the Associate Team Program VERSAMUS}.

\end{thebibliography}
\end{document}